%% file: ccnx.tex
\documentclass[conference]{IEEEtran}
\pdfoutput=1

\usepackage[utf8]{inputenc}
\usepackage[english]{babel}
\usepackage{csquotes}
\usepackage{graphicx}
\usepackage{booktabs}
\usepackage{multirow}
\usepackage{capt-of}

\usepackage[
bookmarks=false, 
colorlinks=true,
linkcolor=black,
citecolor=black,
urlcolor=black
]{hyperref}

\newcommand{\anf}[1]{``#1''}

\IEEEoverridecommandlockouts
\IEEEpubid{\makebox[\columnwidth]{ 978-1-4799-5804-7/15/\$31.00 ©2015 IEEE
 \hfill} \hspace{\columnsep}\makebox[\columnwidth]{ }}

\begin{document}
\input{content/title}
\input{content/abstract}


\input{content/intro}
\input{content/icn_benefits}
\input{content/forwarding_strategies}
\input{content/measurements}

\input{content/ownstrategy}

\input{content/conclusion}

\bibliographystyle{IEEEtran}
\bibliography{content/bib}

\end{document}

%% file: content/title.tex
\title{CCN Forwarding Strategies for \\Multihomed Mobile Terminals}
\date{Faculty Information Systems and Applied Computer Science
University of Bamberg, Germany\\ \bigskip
\today}

\author{\IEEEauthorblockN{Klaus M. Schneider, Kai Mast, Udo R. Krieger}
\IEEEauthorblockA{Otto-Friedrich-University\\
D-96047 Bamberg, Germany\\
Email: klaus.schneider@uni-bamberg.de}
}

\maketitle

%% file: content/abstract.tex
\begin{abstract}
Current IP-based networks are unable to fully exploit the capabilities of the increasing number of multihomed mobile terminals. 
We argue that Content-Centric Networking (CCN), a novel networking architecture based on named information objects, can fill the gap.
In this paper, we elicit requirements for CCN packet forwarding on multihomed mobile terminals. We categorize CCN forwarding strategies according to their ability to fulfill these requirements and provide a real-world performance evaluation in the current CCNx prototype implementation. Moreover, we describe the initial design of an advanced multipath forwarding strategy.
\end{abstract}

%% file: content/intro.tex
\section{Introduction}
\begin{figure*}
\begin{minipage}[t]{.35\linewidth}
\vspace{-7.17em}
\captionof{table}{End-host multihoming in IP}
\label{tab:multihoming_tech}
\footnotesize
\centering
\begin{tabular}{ll}
\toprule
\textbf{TCP/IP layer} & \textbf{Multihoming technology} \\ 
\midrule 
Application & Application-based switching \\ 
\midrule
Transport & SCTP, Multipath TCP \\
\midrule 
Network & Mob. IPv6 \& MCoA, SHIM6 \\
\bottomrule
\end{tabular}
\end{minipage}%
\begin{minipage}[t]{.3\linewidth}
\centering
\includegraphics[width=0.8\linewidth]{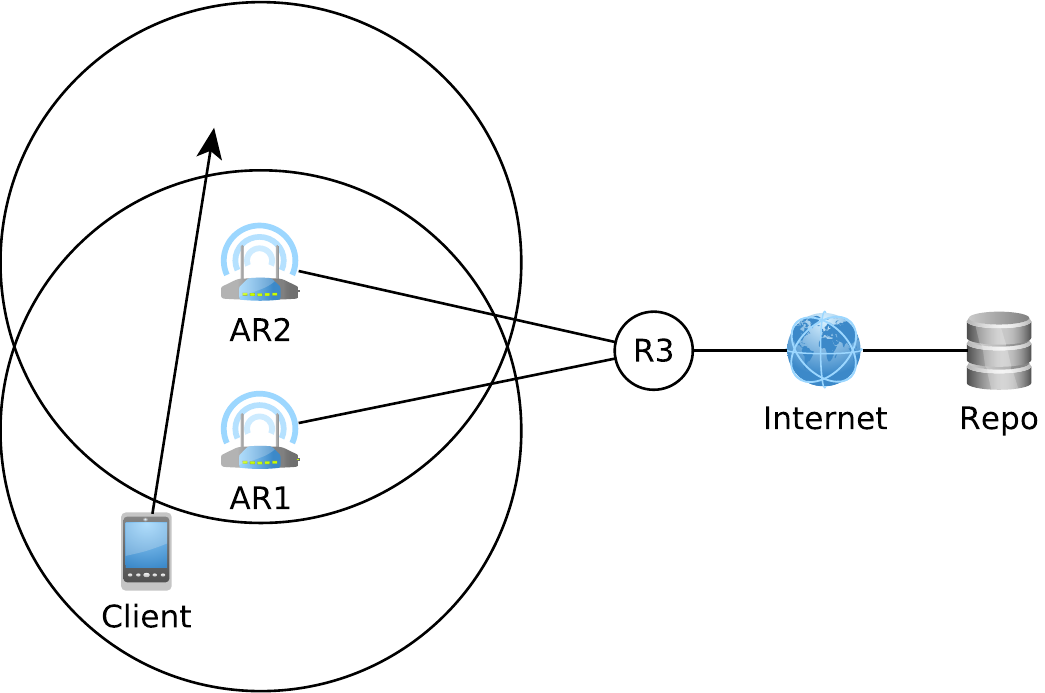}
\caption{CCN terminal handover}
\label{fig:ccn_traveling}
\end{minipage}%
\begin{minipage}[t]{.35\linewidth}
\centering
\includegraphics[width=0.85\linewidth]{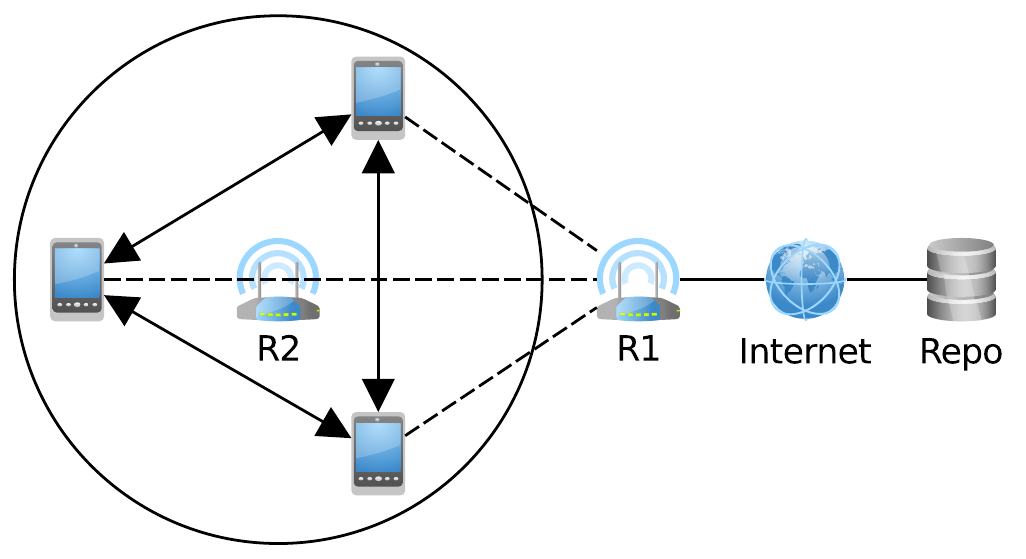}
\caption{P2P content distribution in CCN} 
\label{fig:ccn_p2p}
\end{minipage}
\end{figure*}

Mobile and wireless terminals, e.g. smart phones and tablet pcs, are becoming increasingly popular \cite{cisco_vni}.
These devices are often multihomed, i.e. equipped with multiple wireless network interfaces such as LTE, WiFi and Bluetooth.
Each of these technologies has unique advantages regarding bandwidth, latency, packet loss, security, power consumption, network access costs, etc.
Multiple network interfaces can be combined to exploit the benefits of each technology while simultaneously mitigating its drawbacks. 
However, current IP networks require multiple overlay technologies to perform this task.
IP-based multihoming solutions (Table \ref{tab:multihoming_tech}) have to overcome the \emph{locator/identi\-fier overload}, i.e. IP addresses that are used both as end-point identifiers and locator for routing purposes. 

Most network layer solutions cannot employ network interfaces simultaneously. \emph{IP flow mobility} supports this, but requires the traffic to pass through a redirection infrastructure, e.g. a home agent.
Due to the redirection delay, flow redirection of real-time or delay-sensitive traffic is not considered \cite{ip_flow_mobility}. 
Multipath TCP (MPTCP) is a transport layer protocol that extends TCP by the capability of using multiple simultaneous paths. 
It aims to provide load balancing and app\-li\-cation-trans\-parent fail-over between multiple links, but faces challenges with various types of middleboxes that manipulate the IP header \cite{mptcp_howhard}.
MPTCP was integrated in Apple's iOS 7 as a first wide-scale deployment. Other than that, current mobile operating systems are not able to use multiple network interfaces simultaneously. The most common approach is to switch to the \anf{best} network which is available at a given time, typically WiFi due to lower cost. 

Global Internet bandwidth and the share of content distribution is increasing \cite{cisco_vni} with high-throughput applications such as video streaming and cloud storage. 
Present-day Internet use has shifted from a communication and resource-sharing model (\emph{where}) to a content dissemination model (\emph{what}) \cite{vanJacobson}.
This data-centric paradigm is reflected by today's most popular content distribution mechanisms: Peer-to-Peer (P2P) and Content Distribution Networks (CDNs).

The emerging research field of \emph{Information-Centric Networking} (ICN) addresses this new content-centric usage model by using a clean-slate approach and incorporating named data into the network layer. 
\emph{Content-Centric Networking} (CCN) \cite{vanJacobson}, one of the most promising ICN designs, is intrinsically able to deal with multiple network interfaces and highly fluctuating connectivity. Data retrieval in CCN works connectionless, yet stateful, which avoids issues occurring on disrupted connections. 
CCN's multipath capability has been shown to improve performance in a realistic mobile streaming evaluation using Dynamic Adaptive Streaming over HTTP (DASH) \cite{lederer}. The performance of CCN devices is strongly influenced by its forwarding decisions. However, most current forwarding strategies are very simplistic and designed more as a proof-of-concept than for performance. Moreover, strategies are often designed for CCN core routers \cite{optimal_multipath} rather than terminals. 
In this paper we try to fill this gap by focusing on the classification, measurement-based evaluation and design of forwarding strategies for multihomed terminals.

In the next section we discuss CCN's intrinsic multihoming support. In Section \ref{sec:strategies} we examine requirements of end-device forwarding strategies and discuss specific usage scenarios.  We evaluate the strategies of the current CCNx prototype (Section \ref{sec:evaluation}) and briefly assess the design choices of an advanced forwarding strategy (Section \ref{sec:strategy_design}). 

%% file: content/icn_benefits.tex
\section{Multihoming in CCN} \label{sec:benefits}

CCN has two packet types: \emph{interest} packets to request content objects and \emph{data} packets that contain content chunks. Interests contain a hierarchical name of the requested content and are routed towards a permanent storage location, named \emph{content repository}. 
Data packets travel back the same path of the interests and may be stored temporarily on any intermittent router. 
A content router in CCN consists of three parts: 1) The \emph{Content Store} which caches the content temporarily, 2) the \emph{Pending Interest Table} (PIT) that stores a mapping of a content name to a set of interfaces that requested the content and 3) the \emph{Forwarding Information Base} (FIB) which stores a mapping from content name prefixes to a set of outgoing interfaces (instead of exactly one in IP).
Since interest packets carry a nonce and data packets follow the path of the interests, both CCN packet types are \emph{loop-free}. This feature together with the multipath FIB allows CCN to make intelligent and fine-granular forwarding decisions which are controlled by the \emph{CCN strategy layer}. 
Multihomed terminals benefit from that in a number of ways:

\emph{1) Load balancing.}
CCN can intrinsically combine the bandwidth of multiple network access links. 
The distribution of single interest packets allows a finer control than IP flow distribution.

\emph{2) Support for delay-sensitive applications.} CCN can send interest packets redundantly on multiple links and accept the fastest returning data packet. This concept can improve performance of delay-sensitive applications, e.g. audio- or video-conferencing, in variable mobile and wireless network environments. 

\emph{3) Improved handover performance.}
CCN allows for seamless consumer mobility during handovers between spatially distributed access networks of the same technology (\emph{horizontal handovers}) and different technologies, i.e. \emph{vertical handovers}. In addition, CCN's universal caching improves the end-user performance and reduces the load on the core network as outlined below.
Figure \ref{fig:ccn_traveling} shows a mobile client traveling between two wireless access points. After the handover, all data packets that are still on the path to the client are lost at the broadcast link of \emph{AR1}. In an IP network the client would send the request again, now along \emph{AR2}, and the data would have to travel all the way back from the server. In CCN the client re-issues the interests to \emph{AR2} and most likely finds a copy of the data in the Content Store of a shared core router (\emph{R3}).
Due to the spatial proximity of the two access networks the distance from the Client to R3 is likely significantly smaller than the distance to the repository. Therefore, the response time and the core network load are reduced especially when handovers are frequent.
Tyson et al. \cite{mobility_survey} have stated that mobile terminals can leave a large amount of stale data packets in the network and suggest a redirection mechanism. 
We argue that universal caching mitigates this effect so that data packets will only cause unnecessary traffic load on the path between the shared router and the old position of the client. 
Depending on the relative cost of these links at the network edge, one should consider if the benefits of a redirection mechanism outweigh the benefits of a simpler and more location-agnostic network. 

\emph{4) Ad-hoc P2P content distribution.}
An appropriate CCN forwarding strategy can provide ad-hoc P2P content distribution in local networks (Figure \ref{fig:ccn_p2p}). 
By exploiting its broadcast-capable network interface connected to R2, every CCN terminal can transparently fetch content from local peers instead of the upstream router R1. This is useful in two ways:
a) The upstream link is often more expensive and has lower quality than local broadcast. For instance, LTE typically has higher latency, lower average bandwidth and higher cost than WiFi.
b) The cache size of the upstream CCNx router is limited. Clients can use their storage to contribute to the network and increase the aggregate cache hit rate.
Detti et al. \cite{detti_p2p} have implemented an adaptive video streaming application that exploits an inexpensive proximity network that is used in a P2P fashion together with a more expensive cellular network. 
However, their approach is specific to video streaming using MPEG DASH. By using the right forwarding strategy, we argue that CCN is able to deliver this P2P feature for any overlay  application. 
Since most of the aforementioned benefits require an adaption of the CCN strategy layer we investigate CCN's forwarding concept in more detail.

%% file: content/forwarding_strategies.tex
\section{Forwarding Strategies} 
\label{sec:strategies}

The CCN forwarding strategy decides for each interest packet on which outgoing interface it will be forwarded.  
In most cases, the performance of the mobile terminal is limited by the wireless access network and the impact of the core network is negligible. Therefore, we conclude that the forwarding strategy should be determined by the properties of the access networks and application requirements.

\subsection{Properties of Access Networks}
The characteristics of wireless networks, especially path loss, interference and multipath propagation, lead to a higher bit error rate (BER). The modulation technique of the physical layer is often dynamically adapted to keep the BER at an acceptable level in a trade-off with maximum throughput.
Without an error-correction mechanism packets with bit errors have to be discarded which would lead to high packet loss. Therefore, ARQ and FEC protocols are employed to make a trade-off between packet loss and delay (by re-transmissions) or packet loss and information overhead (by the introduction of redundant parity bits).

In IP networks local re-transmission and redundancy is implemented on the link layer, because
the IP protocol (network layer) does not provide this functionality and transport layer end-to-end re-transmissions produce a much higher overhead. 
CCN works on a hop-to-hop basis and has information which is not available to the link layer, e.g. about caching and multiple paths. 
Thus, we argue that in some cases CCN is the better place to perform re-transmission or to introduce redundancy. The interdependence of the CCN layer with the link layer is crucial for optimal performance and we investigate it more deeply in the next section.

\subsection{Application Requirements} \label{sec:app_requirements}

Applications have different requirements to the Quality of Service (QoS) of the network. For instance, a file download is delay- and loss-tolerant and exploits high bandwidth, while a voice call is delay-sensitive and requires little bandwidth. 
A forwarding strategy should aim to achieve optimal performance with regard to the following parameters which characterize application goals and link properties: 
\emph{1) Throughput.} Some applications, e.g. streaming, have a strong requirement for minimum throughput and also low tolerance to throughput variations. Others like file sharing or email are more tolerant in this regard.
\emph{2) Content response delay.} Many conversational applications are very sensitive to higher packet delays. In audio/video conferences, for example, a delay that significantly exceeds two hundred milliseconds leads to an unacceptable QoS. Therefore, in IP this traffic is often prioritized in specific traffic classes.
\emph{3) Packet loss.} The aforementioned delay-sensitive services are often loss-tolerant, but only up to a certain point. File transfers require that every data bit can be reconstructed correctly. However, since they are not delay-sensitive this can happen through higher layer re-transmissions.
\emph{4) Network access costs. } The network access costs usually scale linearly with the amount of data that is transferred. Since some traffic, e.g. voice conversations, uses relatively little bandwidth and QoS has a high priority, the strategy can use redundancy to improve throughput, delay or packet loss in a trade-off with higher network access costs. 
\emph{5) Privacy \& security.} Some applications like online banking or confidential emails have higher requirements for privacy and security. They should use network interfaces that are better suited in this regard, e.g. prefer a wired point-to-point connection over a public WiFi network. 
\emph{6) Power consumption.} Power consumption still is a big challenge in mobile and wireless environments. Therefore, some users may want to trade-off other requirements and receive data over low-power interfaces.

\subsection{Categorization of Forwarding Strategies}

We identify three broad categories of forwarding strategies that exploit multiple network interfaces (Figure \ref{fig:strategy_triangle}).
Each of them can be optimal to reach different application goals on specific access networks: 

\emph{1) Best Interface First.}
Strategies in the Best Interface First (BIF) category send out interest packets on a pre-selected highest priority interface for a given FIB prefix until a problem like congestion or link failure occurs. Subsequently, the interests will be sent on the next best interface and the order of interfaces will be re-calculated taking the new information about the link problem into account.
The available interfaces can be ranked by performance metrics, e.g. delay or packet loss.
Yi et al. \cite{adaptive_forwarding} have outlined a strategy that falls in this category which uses a coloring scheme for the interface retrieval status, interface probing and Interest NACKs to improve performance.
A simple strategy for the scenario shown in Figure \ref{fig:ccn_p2p} could work like this: The client first broadcasts its interest packets on the inexpensive local link. If the content is not found by local peers, the interests are sent over the more expensive link. 
Due to the proximity of the local peers, the increase in content access delay can be kept relatively small. 

\emph{2) Packet Striping.}
CCN can split interest packets on different network interfaces to maximize the overall throughput at the cost of higher average response time. 
One can use a simple round-robin mechanism to split interest packets in equal parts on the available network interfaces. 
However, since this would likely lead to congestion on one interface and/or under-utilization on others, more sophisticated algorithms have  been proposed. 
Carofiglio et al. \cite{optimal_multipath} have implemented a packet striping algorithm that minimizes the number of pending interests on the most loaded interface.
Sometimes a combination of BIF with packet striping strategies is beneficial.
If an application has a certain bandwidth requirement, a single interface, e.g. WiFi, can be used by default and others, e.g. LTE, can be added if the performance falls below a certain threshold or more bandwidth is requested by the application. 

\emph{3) Parallel and Redundant Transmission.}
Parallel strategies send out interest packets redundantly and, thereby, minimize content response time in a trade-off against higher network access costs.
Since interests cannot loop, CCN can forward them simultaneously on multiple outgoing interfaces, use the first data packet that arrives and discard the ones arriving later. This reduces the overall content response time (VRTT) to the minimum of the response times of all involved interfaces (IF): $ VRTT = min\ \lbrace VRTT_i\ |\ i \in IF \rbrace$.
Note that this sort of redundant traffic is not possible above the link layer in current IP networks, since IP is not loop-free and only supports single-path routing. 

%% file: content/measurements.tex
\section{Experimental Evaluation} \label{sec:measurements} \label{sec:evaluation}

\begin{figure}
\centering
\begin{minipage}[b]{.49\linewidth}
  \centering
  \includegraphics[width=1\linewidth]{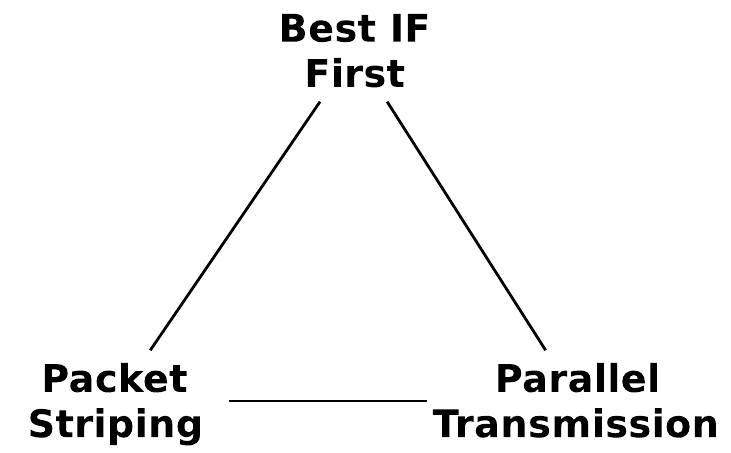}
  \caption{Strategy Triangle}
  \label{fig:strategy_triangle}
\end{minipage}%
\
\begin{minipage}[b]{.49\linewidth}
\centering
\includegraphics[trim=0cm 2mm 0cm 0mm,clip=true,width=0.8\linewidth]{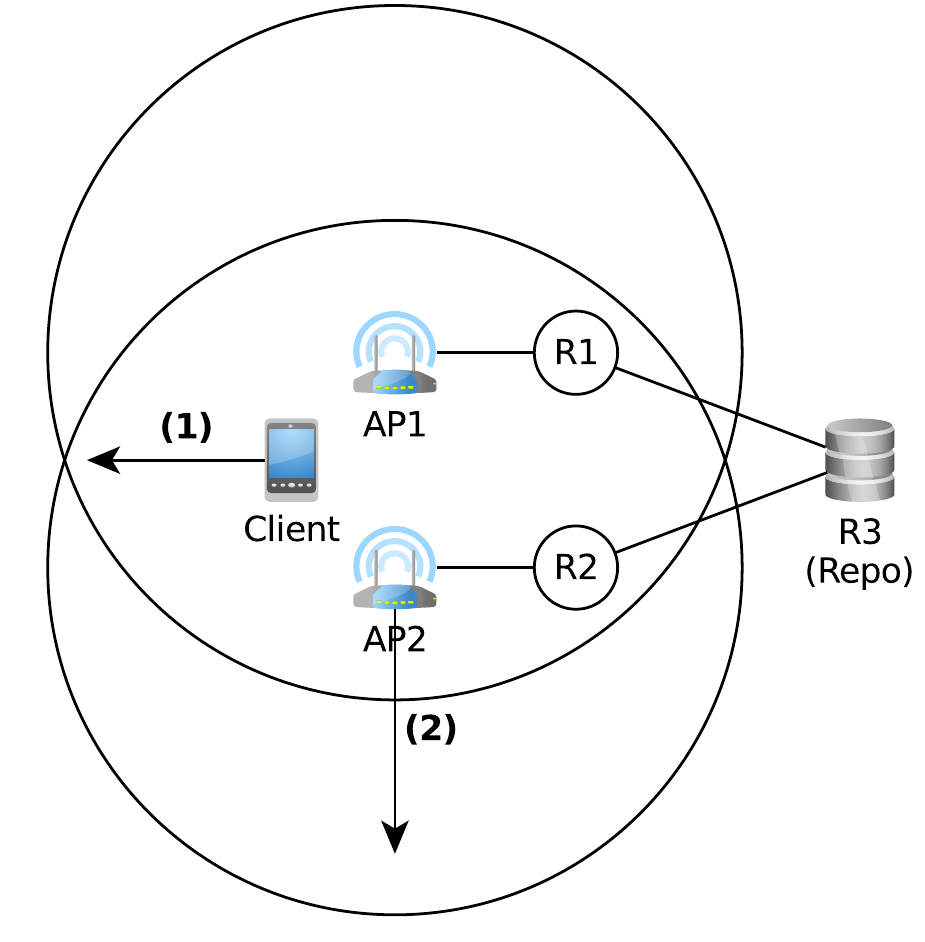}%
\caption{Measurement Topology}
\label{fig:topology}
\end{minipage}
\end{figure}

In this measurement study we want to assess the limitations of current state-of-the-art CCN forwarding strategies in a mobile multihoming scenario. We have compared the strategies that are pre-implemented in the current CCNx version 0.8.2 \cite{ccnx}:
\emph{1) Default.} The default strategy sends interest packets on the fastest responding face. 
After the response time estimate (RTE) has expired it sends them out iteratively (with a random delay) on the other available faces. The RTE is decreased by 1/128 of its current value on every arriving data packet and increased by 1/8 on every timeout. Since this inevitably leads to timeouts about once every 16 packets, the strategy implicitly probes different interfaces to find the best one.
\emph{2) Loadsharing.} The loadsharing strategy sends interest packets on the interface with the smallest number of unanswered interests. This allows a fairly good load balancing and requires very little state to be maintained.
\emph{3) Parallel.} The parallel strategy simply floods interests on all available interfaces.

\subsection{Measurement Setup}

Our measurement topology (Figure \ref{fig:topology}) resembles the assumption that the infrastructure of the core network is negligible. We model the backbone as a black box of a high performance wired network (R3) to isolate the influence of the wireless access network. 
The routers R1, R2 and R3 are desktop PCs with an Intel Core2 E6300 CPU (2x1.86 GHz) and 4 GB Ram. The Client (a laptop with an Intel i5-4300U processor and 8 GB Ram) is equipped with two WiFi AC adapters (Edimax AC-1200). 
Routers and Client are running version 0.8.2 of the CCNx software on Linux. 
The two access points AP1 and AP2 (both Asus RT-AC68U) are configured in the 5 GHz frequency band (max. 867 MBit/s) and bridged to the corresponding router R1 and R2, respectively. 

We noticed that the performance of some CCNx applications, especially those written in Java, was severely limited by the CPU which has been attributed to packet encoding \cite{yuan2011} and heavy state management \cite{wahlisch2011}.
Therefore, we first evaluated the three different file retrieval applications \emph{ccngetfile}, \emph{ccncat} and \emph{ccncatchunks2} (results not depicted) and chose the last one because it showed the best performance. 
ccncatchunks2 features pipelining, i.e. sending out a number of interests in parallel, which, however, is relatively inefficient and sensitive to packet loss due to the early stage of its congestion-control mechanism. Therefore, the pipeline factor should be selected carefully (a value of around 100 was optimal in our case).

\subsection{Reliable Transport}

\begin{figure*}[tb]
\centering
\begin{minipage}[b]{.66\linewidth}
  \centering
  \includegraphics[width=0.49\linewidth]{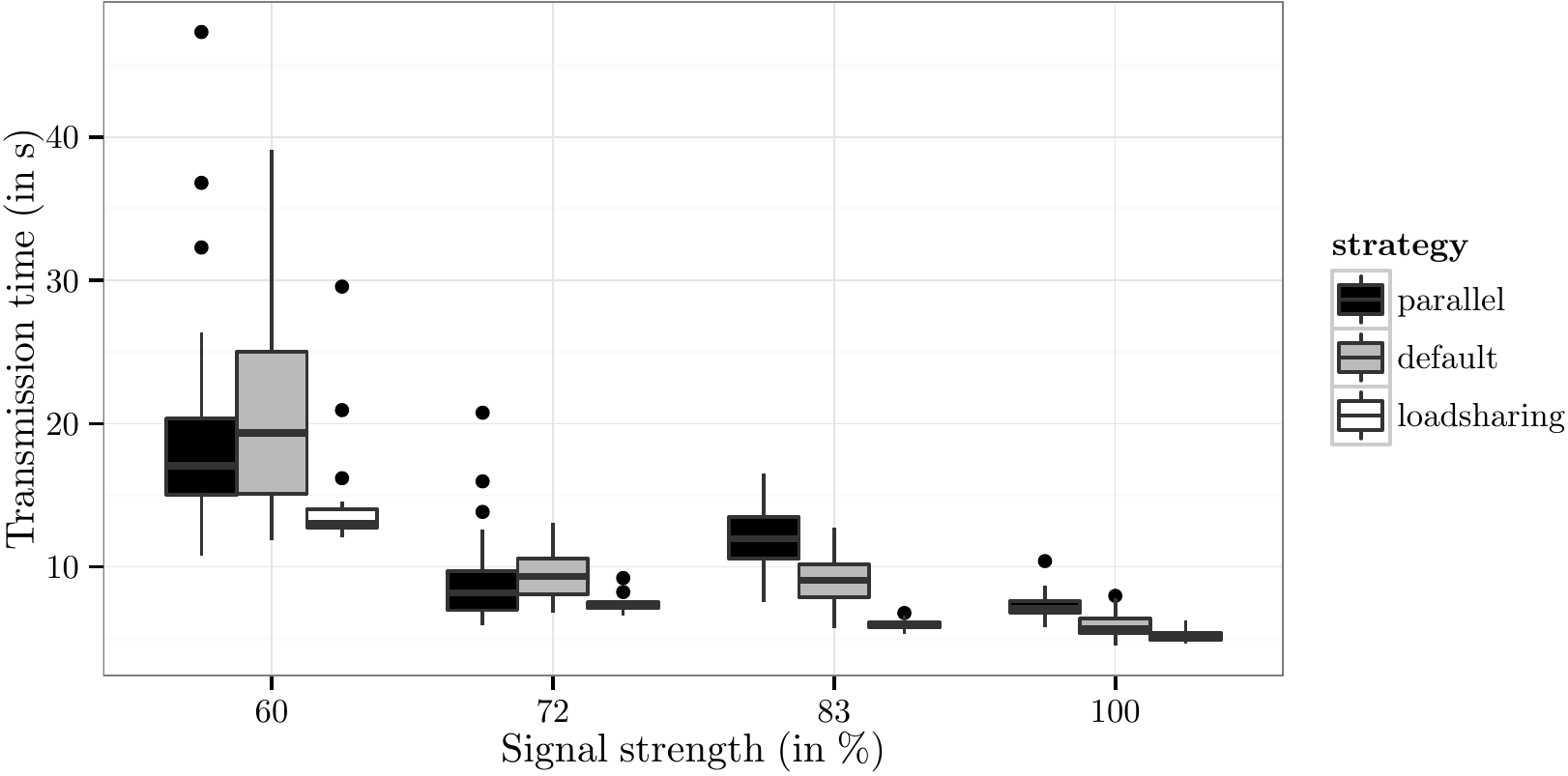}
  \includegraphics[width=0.49\linewidth]{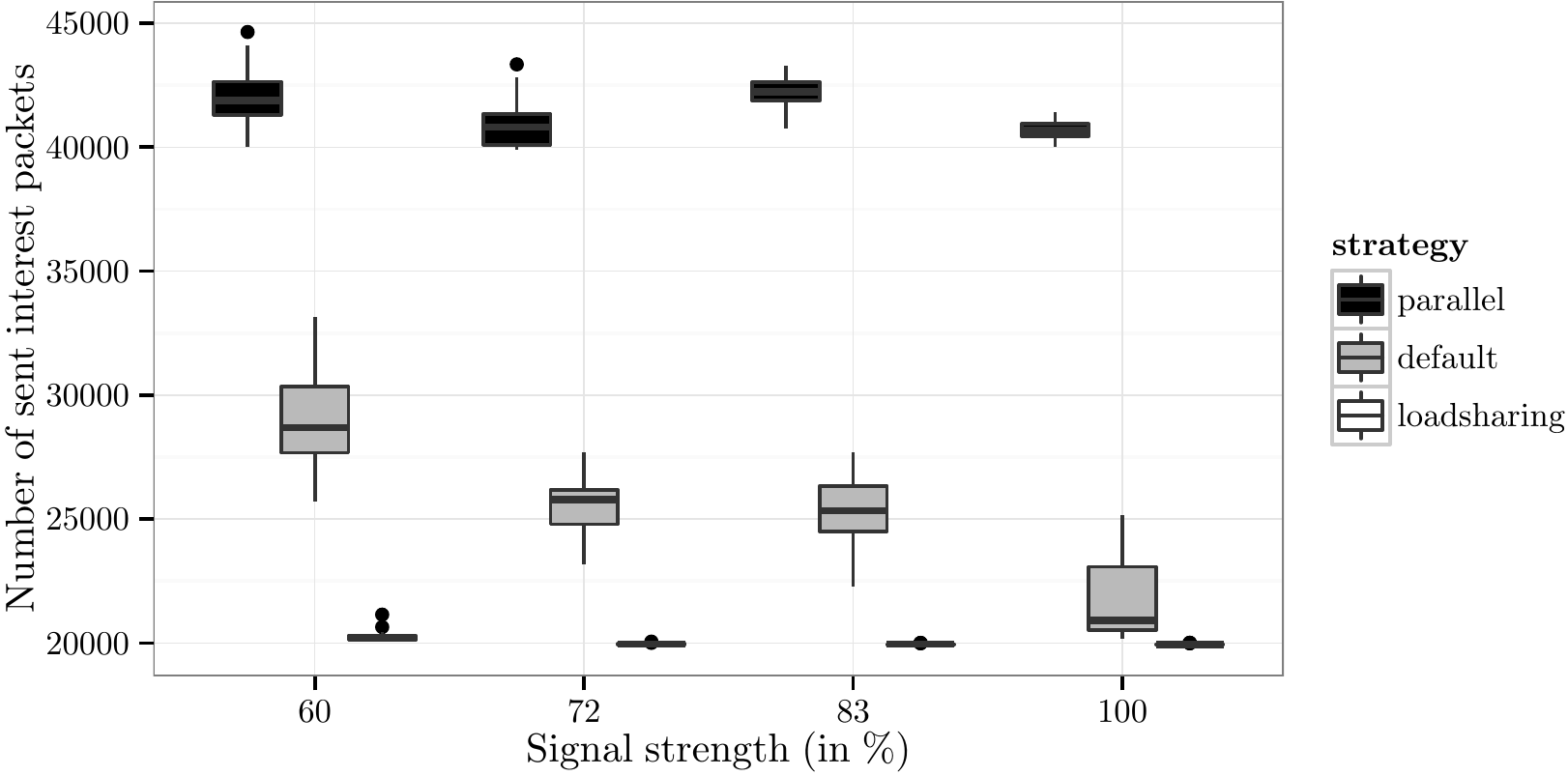}
  \caption{Performance vs. signal quality }
  \label{fig:reliable_mobileclient}
\end{minipage}%
\ 
\begin{minipage}[b]{.33\linewidth}
  \centering
  \includegraphics[width=1\linewidth]{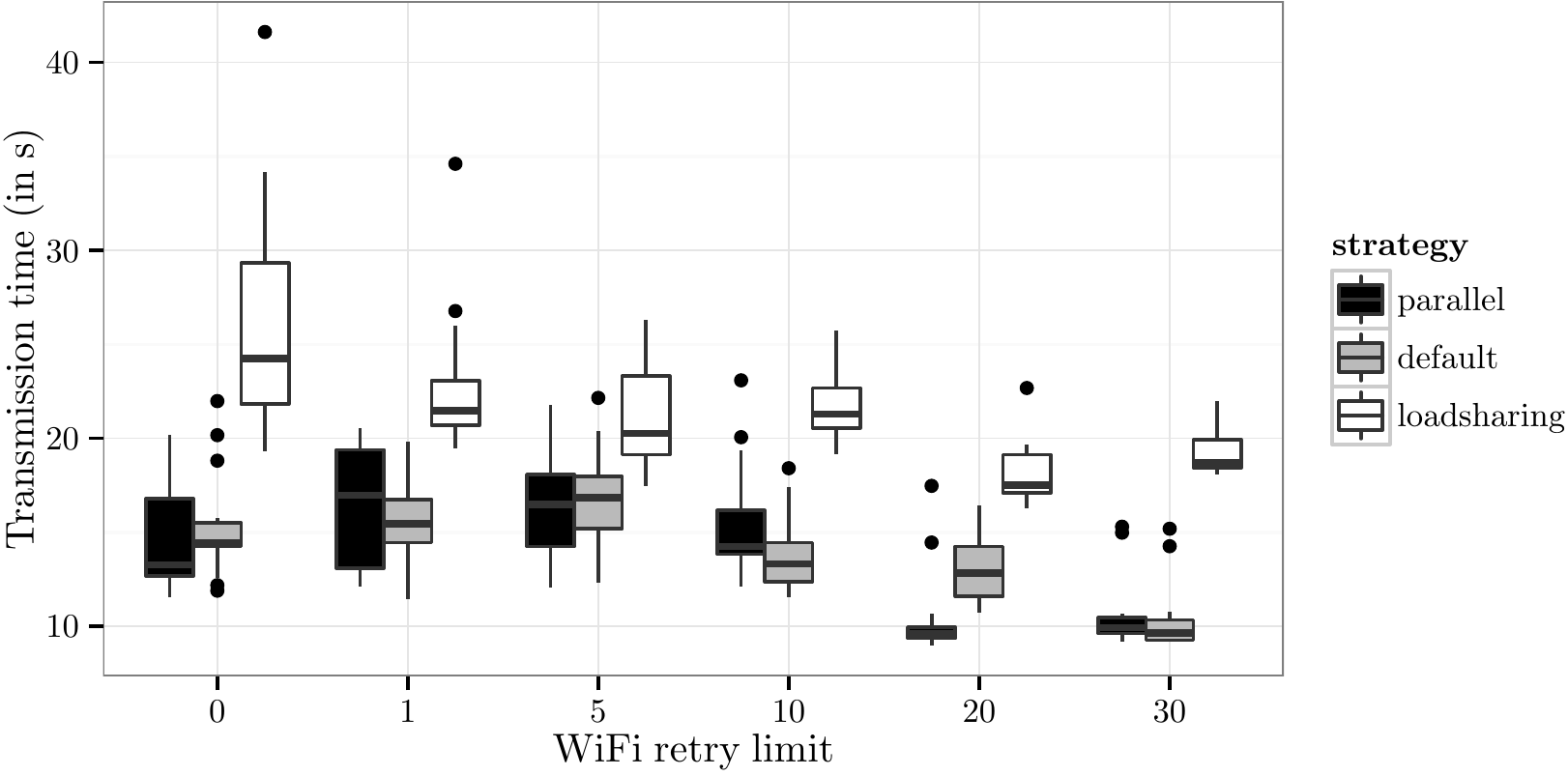}
  \caption{Performance vs. link layer retries}
  \label{fig:retry1}
\end{minipage}%
\end{figure*}

In the first measurement we incrementally moved the client away from both of the two access points to which it was simultaneously connected (Figure \ref{fig:topology} - 1). We evaluated the three  CCNx forwarding strategies using ccncatchunks2 with a pipelining value of 100 transferring 100 MB files with 30 runs each.
We recorded the transmission time and the number of sent interest packets as an estimate of client costs.
Moreover, we recorded the signal strength of the WiFi device (using percentages since the device driver did not provide dB values) and plotted the average of the 30 runs at each measurement location.
In the second measurement we used the same setup, but moved one AP instead of the Client (Figure \ref{fig:topology} - 2). This scenario showed similar, but less pronounced results (not depicted) than the one where both signal qualities varied (Figure \ref{fig:reliable_mobileclient}).

Both measurements show that, as expected, for good reception, the parallel strategy sends roughly twice the amount of interest packets than the other two. While the number of interests sent by the parallel and loadsharing strategy were independent of the signal quality, the default strategy sends more interest packets during poor reception. This is caused by the previously described interface probing mechanism. 
A higher delay and packet loss causes more timeouts on the current interface which makes it more likely that interests are sent on other interface.
Given a good WiFi signal the parallel strategy performed worse than the default strategy. 
This can be explained by the overhead of processing twice the amount of returning data packets. In realistic mobile networks it has to be evaluated if the bottleneck is created by the processing overhead (CPU) or the access links (network).
During poor reception quality the parallel strategy wins ground compared to the others.
We interpret this as evidence that redundancy on the CCN layer can improve performance in this scenario. 

\subsection{Link Layer Re-Transmissions}

In this measurement scenario we want to show the interdependence of the link layer and the CCN strategy layer. On some WiFi cards the tool \texttt{iwconfig} allows the manipulation of the maximum number of link layer re-transmissions per transmitted network layer packet.
Since the Edimax WiFi devices do not support such manipulation (they use a default of 7 retries), we performed this measurement with two cards that do: an on-board Intel Wireless 7260 chip (300 Mbit/s) and one with a Ralink RT3070 chipset (150 Mbit/s). The measurement was performed in a 2,4 GHz WiFi network with good signal quality and 20 runs each.

The results (Figure \ref{fig:retry1}) show that 1)  the loadsharing strategy performed worst in all cases. This may be caused by the fact that the implementation of the loadsharing strategy has problems exploiting channels with the varying performance of two different WiFi cards.
2) At a certain threshold, which is lower for the parallel strategy, a higher retry limit strongly improves the performance. At this point most of the packet loss is handled by the link layer re-transmissions and ccncatchunks2 can work on a loss-free channel. Apparently, re-trans\-missions on the WiFi link layer are at this point much more efficient than those performed by CCNx. 

CCN performance studies normally emulate wireless devices in simulations or testbed experiments with tools like \emph{dummynet} or \emph{tc netem}. This is, to our best knowledge, the first measurement study that evaluates CCN on real wireless devices. Our simple initial design can be extended by using a more complex model of the backbone network, considering caching effects or choosing a greater catalog of transferred files.

%% file: content/ownstrategy.tex
\section{Multipath Strategy Design} \label{sec:strategy_design}

\begin{figure}[tb]
\centering
\includegraphics[width=0.9\linewidth]{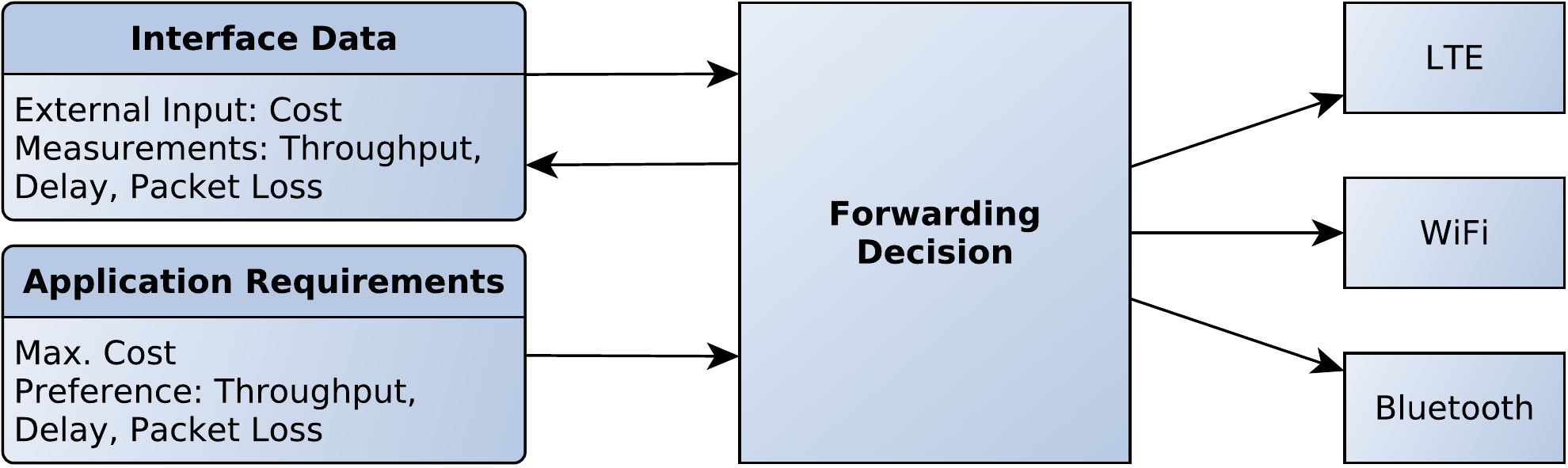}
\caption{High-Level Strategy Architecture} 
\label{fig:own_str}
\end{figure}

The measurement results emphasize the real-world benefit of redundancy on the CCN layer. However, flooding interest packets creates a high overhead and is often unnecessary. Therefore, we propose a forwarding strategy that uses \emph{selective redundancy}, i.e. only sending out interests simultaneously on multiple interfaces if that improves QoS. Moreover, it takes application requirements and network access costs into account.
Since design and implementation are subject to changes we provide a high-level outline (Figure \ref{fig:own_str}) rather than details.  
Each interface is assigned a cost value either by user input or by querying hardware information, e.g. 0.5 for Bluetooth, 1 for WiFi and 3 for LTE. 
Each application can set a maximum cost value and a preference for a specific performance metric. For example, a file download application can choose a maximum cost of 1 and have a preference for maximum throughput. The strategy then chooses the interface that is expected to provide the highest throughput (measured by the amount of returning data packets) while considering the cost constraint. 
An audio conferencing application, on the other hand, can request a delay below 200 ms without a cost limit. The strategy then tries to minimize costs while fulfilling the delay constraint. Most often this will mean sending out interests on one face. However, regarding poor reception quality the strategy will flood the interest packets to achieve the best possible QoS.
This dynamic adaptation provides a considerable advantage over the currently available static forwarding strategies.

After starting the implementation of the strategy in CCNx we switched to the NDN Platform \cite{ndn_platform}. The NDN Forwarding Daemon (NFD) is built on modular C++ code and open to community contributions. The design of our solution is portable to other ICN architectures as long as they feature multipath routing and a stateful forwarding plane. 
The implementation requires several iterations of testing and performance evaluation to become mature.
Thereafter, we plan to release it under an open source license.

%% file: content/conclusion.tex
\section{Conclusion}  \label{sec:conclusion}

CCN's stateful forwarding plane creates many opportunities to overcome current challenges of multihomed terminals.
A CCN forwarding strategy can perform local re-transmissions and exploit knowledge of multiple paths on the network layer. This leads to the novel question of how to divide the responsibilities between the CCN network layer and the link layer. Sophisticated forwarding strategies are possible that are tailored to application requirements and can use the available access networks in a more efficient way. However, current state-of-the-art strategies are still in very early stages
and more research is needed in designing, implementing and evaluating better alternatives.